\documentclass[10pt,onecolumn]{article}

\usepackage{graphicx}
\usepackage{amsfonts}
\usepackage{amsmath}
\usepackage{amssymb}
\usepackage{mathrsfs} 
\usepackage{lipsum}
\usepackage{color}
\usepackage{siunitx}
\usepackage{authblk}
\usepackage[margin=2cm]{geometry}
\usepackage{mathtools, cuted}
\usepackage{breqn}

\usepackage{bm}
\usepackage{amsmath}

\title{The break-up of free films pulled out of a pure liquid bath}

\author[1]{Lor\`{e}ne Champougny}
\author[1]{Emmanuelle Rio} 
\author[1]{Fr\'{e}d\'{e}ric Restagno}
\author[2]{Benoit Scheid}

\affil[1] {Universit\'e Paris-Sud, Laboratoire de Physique des Solides, UMR8502, Orsay, F-91405}
\affil[2] {TIPs - Fluid Physics Unit, Universit\'{e} Libre de Bruxelles C.P. 165/67, 1050 Brussels, Belgium}

\date{\today}%

\newcommand{\arctanh}{\mathrm{arctanh}}
\newcommand{\eps}{\varepsilon}
\newcommand{\mms}{\mathrm{mm/s}}
\newcommand{\mNparm}{\mathrm{mN/m}}
\newcommand{\kilogrampercubicmetre}{\text{kg.m$^{-3}$}}
\newcommand{\Ca}{\mathrm{Ca}}
\newcommand{\We}{\mathrm{We}}
\newcommand{\uu}{\bar{u}}
\newcommand{\hstar}{h^{\ast}}
\newcommand{\Lstar}{L^{\ast}}
\newcommand{\Heff}{H_{\mathrm{eff}}}
\newcommand{\dd}{\mathrm{d}}
\newcommand{\Aprime}{\mathcal{A}^{\prime}}
\newcommand{\joule}{\mathrm{J}}
\newcommand{\milli}{\mathrm{m}}
\newcommand{\centi}{\mathrm{c}}
\newcommand{\meter}{\mathrm{m}}
\newcommand{\per}{\mathrm{/}}
\newcommand{\second}{\mathrm{s}}
\newcommand{\pascalsecond}{\mathrm{Pa.s}}
\newcommand{\micro}{\mu}
\newcommand{\kilogram}{\mathrm{kg}}
\newcommand{\mole}{\mathrm{mol}}
\newcommand{\celsius}{\deg\mathrm{c}}
\newcommand{\gram}{\mathrm{g}}
\newcommand{\hour}{\mathrm{h}}

\begin{document}
\maketitle
%
%
\begin{abstract}
In this paper, we derive a lubrication model to describe the non-stationary free liquid film that is created when a vertical frame is pulled out of a liquid reservoir at a given velocity. We here focus on the case of a pure liquid, corresponding to a stress-free boundary condition at the liquid/air interfaces of the film, and thus employ an essentially extensional description of the flow. Taking into account van der Waals interactions between the interfaces, we observe that film rupture is well-defined in time as well as in space, which allows us to compute the critical thickness and the film height at the moment of rupture. The theoretical predictions of the model turn out to be in quantitative agreement with experimental measurements of the break-up height of silicone oil films in a wide range of pulling velocities and supporting fiber diameters.
\end{abstract}
%
%
\section{Introduction}
%

Everyday life experience shows that thin films or bubbles made of pure liquids -- such as oil or pure water -- are usually short-lived, while aqueous solutions containing surface active molecules can give rise to much more stable films. Yet, even in the absence of surfactant, gas entrapment in highly viscous liquids can lead to the formation of bubble layers, for example at the surface of molten glass in furnaces \cite{Kappel1987} or in lava flows \cite{Proussevitch1993}. The stability of such structures has been studied by \cite{Debregeas1998} and then by \cite{Kocarkova2013} through drainage measurements on individual viscous bubbles floating at the surface of a liquid pool. A different film geometry, in which a vertical free film is pulled out of a soapy solution, has been widely investigated in both experiments \cite{Mysels1959, vanNierop2008, vanNierop2009_formation, Saulnier2011} and \emph{stationary} models \cite{Mysels1959, Seiwert2014, Champougny2015}. However, the pulling of viscous surfactant-free films, in which the flow is intrinsically non-stationary, has attracted much less attention so far.

A fundamental difference between pulling a soap film and a pure liquid film out of a bath lies in the nature of the induced flow. In the absence of surfactants, the film liquid/air interfaces are stress-free, leading to an extensional flow with a uniform velocity in the direction transverse to the film \cite{Howell1996}. Extensional film flow corresponds mathematically to a distinguished limit, as described by \cite{Breward_phd}, in which extensional viscous stresses balance all other forces along the film such as inertia, surface tension, gravity and van der Waals forces. The addition of surfactant molecules usually gives rise to a non-zero tangential stress at the interfaces, for example due to surface tension gradients induced by surfactant concentration gradients, and the flow profile has a parabolic shape \cite{Mysels1959}. \cite{Scheid2012_thermocapillary} have shown that sufficient shear stress can also be obtained by imposing a sharp temperature gradient near the liquid bath such that a stationary film of pure liquid can still be formed. In the present paper however, we only consider films with stress-free interfaces such that the flow is unsteady and of extensional nature. 

Lubrication models are extensively used in literature to describe free-film time evolution, which can be split into a stretching phase and a break-up phase, the latter occurring on a much shorter timescale that the former. For instance, \cite{Erneux1993} and \cite{Vaynblat2001} describe the mechanism of free-film rupture for unbounded films and both highlight the importance of nonlinear contributions to the acceleration of the rupture phenomenon. Similarly, \cite{Tabakova2010} have analysed the stability of free films attached to lateral plates by a given contact angle, the value of which determines whether static solutions are unstable or not under asymmetrical perturbations. However, all these works assume a uniform base state, \textit{i.e.} an initial film of constant thickness and zero velocity, which is then perturbed. Hence, they do not capture the influence of the pulling dynamics on film rupture. To our knowledge, the pull-up of a two-dimensional film above a liquid bath has only been reported by \cite{PhD_Heller}. Using time-dependent simulations of the two-dimensional Stokes equations with a moving mesh, \cite{PhD_Heller} was able to simulate the stretching of a viscous liquid film over a wide range of pulling velocities, showing that the length of the film increases with the pulling velocity. 

In a different geometry, the pulling and break-up of axisymmetric liquid bridges have recently attracted much attention, starting with the work of \cite{Marmottant2004} who studied the stretching of liquid ligaments at small and large extension rates, and investigated the fragmentation scenario in the last case. For a liquid bridge between a bath and a perfectly wetting horizontal disk above, \cite{Benilov2010} reported that the maximum height of the liquid bridge in the static limit is twice the capillary length, denoted $\ell_c = \sqrt{\gamma/\rho g}$, where $\gamma$ is the surface tension, $\rho$ is the liquid density and $g$ is the gravitational acceleration. This maximum height was furthermore shown by \cite{Benilov2010} to decrease with increasing contact angle on the disk. Later, \cite{Benilov2013} showed that the stability of their liquid-column solutions strongly depends on the dynamics of the contact line, and in more general terms on the nature of the boundary condition at the disk. For a liquid bridge between two parallel horizontal disks, \cite{Chen2015} have explored the influence of surface wettability on the transition from a quasi-static (capillary dominated) to a dynamic stretching (dominated by viscous and inertial forces). On the same system, experiments and simulations have been performed to understand the influence of the stretching velocity \cite{Zhuang2015} or acceleration \cite{Weickgenannt2015} on the time of liquid bridge break-up. 

In this work, we present a non-stationary lubrication model describing the thin liquid film that is formed when pulling a horizontal fiber from pure liquid bath, starting with an initially static meniscus. The lubrication model is presented in section~\ref{sec:model} and validated in section~\ref{sec:validation} using the two-dimensional simulations of \cite{PhD_Heller}. Results including inertia and van der Waals interactions are next shown in section~\ref{sec:results_model}. An experimental study is presented in section~\ref{sec:exp}, whose results are shown and compared to simulations in section~\ref{sec:comparison}. Conclusions are given in section~\ref{sec:conclusions}.
%
\section{Lubrication model for vertical film pulling} \label{sec:model}
%
\subsection{Problem settings}
%
\begin{figure}
\centering
\includegraphics[width=\linewidth]{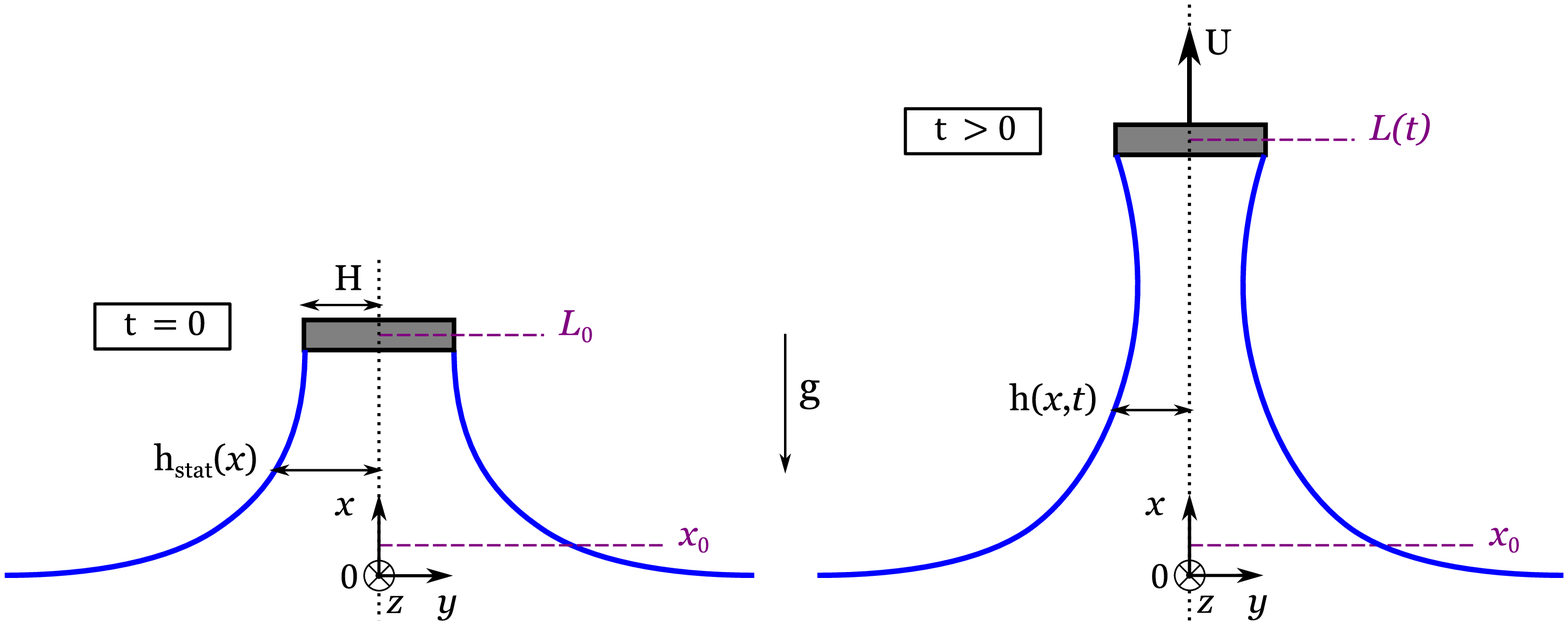}
\caption{Sketches of the liquid film under consideration, showing the notations and in particular the locations $x_0$ and $L(t)$ where the boundary conditions are imposed. The initial configuration is presented on the left panel, where a static meniscus connects the fiber, at the initial position $L_0$, to the liquid bath located at $x=0$. The right panel represents the liquid film at a later time, during pulling at a constant velocity $U$.}
\label{fig:notations}
\end{figure}
We seek to describe the thin liquid film that is created when a perfectly wetting horizontal fiber is vertically pulled out of a liquid bath at a constant velocity $U$ (in the frame of the laboratory). The liquid under consideration is incompressible and Newtonian, of dynamic viscosity $\eta$, density $\rho$, and surface tension $\gamma$. As pictured in figure \ref{fig:notations}, the fiber is parallel to the $z$-axis and invariant by translation along this axis. It is supposed to have a rectangular cross-section, of width $2H$, such that the liquid film is pinned at $y=\pm H$. The film is assumed to be symmetric with respect to the vertical $x$-axis, so that only the half film thickness $h(x,t)$ will be considered in the following.

It is also assumed that the film is uniform in the lateral direction, \textit{i.e.} perpendicular to the $xy$-plane, such that edge effects are ruled out. During film pulling, the vertical and horizontal velocity fields, respectively denoted $u(x,y,t)$ and $v(x,y,t)$, and the pressure field $P(x,y,t)$ obey the following governing equations and boundary conditions:
\begin{itemize}
\item Navier-Stokes (momentum) equations:
	\begin{subequations}	\label{eq:NS}
	\begin{alignat}{1}
	\rho \, (\partial_t u + u \partial_x u + v \partial_y u) &= -\partial_x P +\eta \, (\partial_{xx}u + \partial_{yy}u) -\rho g\,, \label{eq:NSx}\\
	\rho \, (\partial_t v + u \partial_x v + v \partial_y v) &= -\partial_y P +\eta \, (\partial_{xx}v + \partial_{yy}v)\,, \label{eq:NSy}
	\end{alignat}
	\end{subequations}
\item continuity equation:
	\begin{equation} 	
	\partial_x u + \partial_y v = 0\,, \label{eq:continuity}
	\end{equation}
\item symmetry conditions at $y=0$:
        \begin{subequations} \label{eq:symmetry}
	\begin{alignat}{1}
	v &= 0\,, \label{eq:symmetrya}\\
	\partial_y u &=0\,,\label{eq:symmetryb} 
	\end{alignat}
	\end{subequations}
\item kinematic condition at $y=h(x,t)$:
	\begin{equation} 	
        v  = \partial_t h + u\partial_x h\,,\label{eq:kinematic}
	\end{equation}
\item tangential and normal stress balances at $y=h(x,t)$:
	\begin{subequations} \label{eq:surf_balance}
	\begin{alignat}{1}
	 \frac{ \eta}{1 + (\partial_x h)^2} \left[ \left( 1-(\partial_x h)^2 \right) ( \partial_y u + \partial_x v )- 4 \partial_x h \partial_x u  \right] &= 0\,, \label{eq:surf_balance_T}\\
	 P_0 - P - \frac{2 \eta}{1 + (\partial_x h)^2} \left[ \left( 1-(\partial_x h)^2 \right) \partial_x u + \partial_x h ( \partial_y u + \partial_x v ) \right]  &= 2 \gamma K\,, \label{eq:surf_balance_N}
	\end{alignat}
	\end{subequations}
in which the continuity equation \eqref{eq:continuity} has been used, and where $P_0$ is the atmospheric pressure and $K$ is the mean curvature of the interface expressed as
	\begin{equation}
	K(x,t) = \frac{1}{2} \frac{\partial_{xx}h}{\left[1 + (\partial_x h)^2 \right]^{3/2}}\,.
	\label{eq:curvature}
	\end{equation}
\end{itemize}
The continuity equation \eqref{eq:continuity} can then be integrated along the $y$-coordinate, using the Leibniz integral rule and the boundary conditions \eqref{eq:symmetrya} and \eqref{eq:kinematic}, such that it yields the mass conservation equation,
	\begin{equation}
	\partial_t h + \partial_x (\bar{u} h) = 0\,,
	\label{eq:mass_cons_approx}
	\end{equation}
where
	\begin{equation}
	\bar{u}(x,t) = \frac{1}{h} \int_0^{h} \!u \, \mathrm{d}y
	\label{eq:vitesse_moy}
	\end{equation}
is the average velocity in the film at a given height $x$ and time $t$.
%
\subsection{Lubrication approximation and velocity decomposition\label{scalings_lubrication}}
%
We first pose that the characteristic lengths in the $x$- and $y$-directions are, respectively, the capillary length $\ell_c = \sqrt{\gamma / \rho g}$ and the fiber half-width $H$, such that the aspect ratio $\eps$ is defined as
	\begin{equation}
	\eps = \frac{H}{\ell_c} \ll 1.
	\label{eq:epsilon}
	\end{equation}
The smallness of $\eps$ corresponds to small space modulations of the dependent variables inherent to the lubrication approximation.

Next, it is well established (see e.g. \cite{Howell1996}) that when no shear is applied at the interfaces -- \textit{i.e.} the right-hand-side of \eqref{eq:surf_balance_T} is zero --, the flow at leading order is purely extensional (\textit{i.e.} a plug flow) and that the vertical velocity is uniform and equal the average velocity $\uu$. Therefore, the vertical velocity field is split into a plug flow contribution $\uu(x,t)$ and a shear contribution $u_1(x,y,t)$ as
	\begin{equation}
	u(x,y,t) = \uu(x,t) + u_1(x,y,t),
	\label{eq:decompose_u}
	\end{equation}
where the dominant contribution $\uu$ scales as $U$, while the shear contribution $u_1(x,y,t)$ is assumed to be of order $\eps^2 U$. Conservation imposes that
	\begin{equation}
	\int_0^{h} u_1 \, \mathrm{d}y = 0.
	\label{eq:zero_flux}
	\end{equation}
A similar velocity decomposition has been proposed by \cite{Schwartz1999} aiming at modelling surfactant-stabilised liquid films. For this purpose, they rather imposed a no-slip condition $u_1(x,h(x,t),t)=0$ for the shear correction, such that the leading order velocity contribution coincides with the interfacial velocity $u_s(x,t)$, instead of $\uu(x,t)$ in the present case. The justification of Schwartz \& Roy's approach relies in the fact that $u_s$ explicitly appears in the transport equation of surfactant at the interface, while such equation is absent in the case of pure liquids.

The continuity equation imposes the horizontal velocity $v$ to be of order $\eps U$, the pressure $P$ is scaled by $\eta \, U / \ell_c$ and the time $t$ by $\ell_c / U$. The system of equations~(\ref{eq:NS}--\ref{eq:surf_balance}), in which \eqref{eq:epsilon} and \eqref{eq:decompose_u} are used, is then non-dimensionlised and truncated at $\mathcal{O}(\eps^2)$. Making the resulting equations dimensional again yields the following approximated system:
\begin{itemize}
\item momentum equations:
	\begin{subequations}	\label{eq:NS_approx}
	\begin{alignat}{1}
	\eta \,\partial_{yy}u_1 &= \partial_x P  + \rho g + \rho \, (\partial_t \uu + \uu \partial_x \uu) - \eta \,\partial_{xx}\uu \,, \label{eq:NSx_approx} \\
	\partial_y P &= 0\,, \label{eq:NSy_approx}
	\end{alignat}
	\end{subequations}
\item continuity equation:
	\begin{equation} 	
	\partial_x \uu + \partial_y v = 0\,, \label{eq:continuity_approx}
	\end{equation}
\item symmetry conditions at $y=0$:
    \begin{subequations} \label{eq:symmetry_approx}
	\begin{alignat}{1}
	v &= 0\,, \label{eq:symmetry_approxa}\\
	\partial_y u_1 &=0\,,\label{eq:symmetry_approxb} 
	\end{alignat}
	\end{subequations}
\item tangential and normal stress balances at $y=h(x,t)$:
	\begin{subequations}	\label{eq:surf_balance_approx}
	\begin{alignat}{1}
	\partial_y u_1 &= 4 \partial_x h \, \partial_x \uu - \partial_x v \,,  \label{eq:surf_balance_approxa}\\
	P &= P_0 - 2\eta \, \partial_x \uu  - 2 \gamma K  \,. \label{eq:surf_balance_approxb}
	\end{alignat}
	\end{subequations}
\end{itemize}
The above system is closed by the unchanged conservation equation~\eqref{eq:mass_cons_approx}. Note that we did not truncate at $\mathcal{O}(\eps^2)$ the expression \eqref{eq:curvature} for the mean curvature, despite the higher-order correction at the denominator, as done for liquid bridges by \cite{Vincent2014} for example. This keeps the model compatible with the solution of the static meniscus near the bath, where the slope is not small anymore, and allows describing the entire film shape, from the bath up to the fiber, without having to use a matching procedure between the film region and the static meniscus.

Integrating \eqref{eq:NSy_approx} leads to a constant pressure in the horizontal direction, whose expression is then given by \eqref{eq:surf_balance_approxb}. 
In order now to account for van der Waals interactions leading to film rupture, we use the concept of disjoining pressure introduced by \cite{Derjaguin1936}. These authors stated that the pressure in a thin film is modified by the molecular interactions as compared to the surrounding bulk phase such that \eqref{eq:surf_balance_approxb} is rewritten as follows
	\begin{equation} 	
	P = P_0 - 2\eta \, \partial_x \uu  - 2 \gamma K - \Pi, 
	\label{eq:surf_balance_approxb_PI}
	\end{equation}
where $\Pi(x,t)$ is the disjoining pressure. Classical theory predicts that the attractive London-van der Waals interaction for a film of thickness $2h$ can be expressed as \cite{Israelachvili2011}
	\begin{equation}
	\Pi(x,t) = -\frac{A_H}{6 \pi (2h)^3}, 
	\label{eq:vdW}
	\end{equation}
where $A_H>0$ is the non-retarded Hamaker constant. Note that even though intermolecular forces are bulk contributions, they are equivalently introduced as surface forces in the frame of the lubrication approximation, as thoroughly discussed by \cite{Ivanov1988}.

Finally, using \eqref{eq:continuity_approx} and~\eqref{eq:symmetry_approxa}, the horizontal velocity can be expressed as 
	\begin{equation}
	v(x,y,t) = - \partial_x \uu \, y, 
	\label{eq:v_approx}
	\end{equation}
and the shear component of the vertical velocity $u_1$ is obtained by integration of \eqref{eq:NSx_approx} using \eqref{eq:zero_flux} and~\eqref{eq:symmetry_approxb},
	\begin{equation}
	u_1(x,y,t) = \frac{1}{2\eta} \left[\partial_x P  + \rho g + \rho \, (\partial_t \uu + \uu \partial_x \uu) - \eta \,\partial_{xx}\uu  \right] \left(y^2 - 	\frac{h^2}{3} \right).
	\label{eq:u1_profile}
	\end{equation}
Substituting \eqref{eq:v_approx} and \eqref{eq:u1_profile} into \eqref{eq:surf_balance_approxa}, while using \eqref{eq:surf_balance_approxb_PI} and \eqref{eq:vdW}, finally yields
	\begin{equation}
	\rho h (\partial_t \uu + \uu \partial_x \uu) - h\left(2\gamma \partial_x K - \rho g + \frac{A_H}{16\pi}\frac{\partial_x h}{h^4}  \right) - 4\eta \, \partial_{x}(h \partial_x \uu) = 0.
	\label{eq:tang_stress_dim}
	\end{equation}
The dimensional system of equations to be solved consists of equations \eqref{eq:curvature}, \eqref{eq:mass_cons_approx} and~\eqref{eq:tang_stress_dim}, for the three unknowns, $K$, $h$ and $\uu$.
%
\subsection{Non-dimensionalised problem} \label{sec:ndim_problem}
%
Applying the following transformations,
	\begin{equation}
	h \rightarrow H h, \quad x \rightarrow \frac{H}{\eps} x, \quad \uu \rightarrow U \uu, \quad t \rightarrow \frac{H}{\eps U} t \quad\text{and}\quad K		\rightarrow \frac{\eps^2}{H} K, 
	\label{eq:scaling}
	\end{equation}
equations~\eqref{eq:mass_cons_approx} and~\eqref{eq:tang_stress_dim} become in dimensionless form
	\begin{subequations}	\label{eq:system_ndim}
	\begin{alignat}{1}
	\partial_t h + \partial_x \left( \uu h \right) &= 0,
	\label{eq:mass_cons_ndim}\\
	\We \,h \left(\partial_t \uu + \uu \partial_x \uu \right) - h \left(2 \eps \partial_x K - 1 + \mathcal{A} \, \frac{\partial_x h}{h^4} \right) - 4 \Ca \, 		\partial_{x}(h \partial_x \uu) &= 0,
	\label{eq:tang_stress_ndim}
	\end{alignat}
	\end{subequations}
where the dimensionless mean curvature \eqref{eq:curvature} is rewritten as
	\begin{equation}
	K(x,t) = \frac{\partial_{xx}h}{2\left[1 + (\eps \partial_x h)^2 \right]^{3/2}}.
	\label{eq:curvature_ndim}
	\end{equation}
In addition to $\eps$, the dimensionless numbers governing the above system of equations are the capillary number, the Weber number and the Hamaker number, respectively defined as
	\begin{equation}
	\Ca = \frac{\eta \, U}{\gamma}, \qquad \We = \frac{\rho \, U^2 \ell_c}{\gamma} \qquad \text{and} \qquad \mathcal{A} = \frac{A_H \ell_c}{16 \pi \gamma H^3}.
	\label{eq:ndim_numbers}
	\end{equation}
The first two terms in \eqref{eq:tang_stress_ndim} account for inertia while the others account, respectively, for capillary pressure, gravity, intermolecular forces and extensional viscous stress. \bigskip

The dimensional physico-chemical parameters we implement in the simulation are those of the V1000 silicone oil that will be used in the experiments presented in section~\ref{sec:exp}, namely $\eta=1.00~\pascalsecond$, $\rho=970~\kilogrampercubicmetre$, $\gamma=21.1~\mNparm$ and $A_H=4.4\times 10^{-20}~\joule$, yielding a capillary length $\ell_c = 1.49~\milli\meter$. For a given liquid, the physical parameters that can be changed experimentally are the pulling velocity $U$ and the fiber half-width $H$. In the simulation, these parameters will be explored in the ranges $0.001~\milli\meter\per\second \leqslant U \leqslant 30~\milli\meter\per\second$ and $0.1~\micro\meter \leqslant H \leqslant 1000~\micro\meter$, respectively. Consequently, the dimensionless numbers defined in \eqref{eq:ndim_numbers} will vary in the following ranges:
	\begin{subequations} \label{eq:ODG}
	\begin{alignat}{1}
	5\times 10^{-5}&\leqslant \Ca \leqslant 1, \\
	7\times 10^{-11} &\leqslant \We \leqslant 0.06,\\
	8\times 10^{-14}&\leqslant \mathcal{A} \leqslant 0.02, \\
	1\times 10^{-4}&\leqslant \eps \leqslant 0.6.
	\end{alignat}
	\end{subequations}
It is worth noting that the Hamaker number $\mathcal{A}$ can be expressed as a function of $\eps$ as
\begin{equation}
\mathcal{A} = \frac{\mathcal{A}^{\prime}}{\eps^3} \qquad \text{where} \quad \mathcal{A}^{\prime} = \frac{A_H}{16 \pi \gamma \ell_c^2},
\label{eq:nombre_Hamaker_reformulé}
\end{equation}
which we call the dimensionless Hamaker constant, depends solely on the properties of the liquid. Similarly, the Weber number $\We$ can be expressed as a function of the capillary number $\Ca$ and of the liquid properties. Hence, for a given liquid, only $\Ca$ and $\eps$ can be varied independently.
%
%
\section{Model resolution and validation \label{sec:validation}}
%
\subsection{Initial solutions and boundary conditions\label{CI_and_CL}}
%
We consider the initial configuration pictured in the left panel of figure~\ref{fig:notations}, where the fiber is located at a (dimensionless) position $L_0$ above the surface of the liquid bath such that it forms a static liquid bridge. \cite{PhD_Heller} has shown that for $L_0 \leqslant 2$, a solution exists, whose shape results from a balance between the capillary pressure gradient $\gamma \partial_x K$ and the hydrostatic pressure gradient $\rho g$. Integrating the corresponding dimensionless balance
	\begin{equation}
	\partial_x K_{\rm stat} = \frac{1}{2 \eps},
	\label{eq:static_meniscus_ndim}
	\end{equation}
with the boundary conditions $K_{\rm stat}(0) = 0$, $h'_{\rm stat}(0) = -\infty$ and $h_{\mathrm{stat}}(L_0) = 1$, yields an analytical expression for the shape of the static meniscus,
	\begin{equation}
	h_{\mathrm{stat}}(x) = 1 + \frac{1}{\eps} \left[ \sqrt{4-L_0^2} -  \sqrt{4-x^2} - \arctanh \!\left( \frac{2}{\sqrt{4-L_0^2}} \right)  + \arctanh \!\left( \frac{2}{\sqrt{4-x^2}} \right) \right],
	\label{eq:static_meniscus_profile}
	\end{equation}
which is valid in the range $0 < L_0 \leqslant 2$.

The dimensionless initial solutions corresponding to the static meniscus are therefore
	\begin{subequations}	\label{eq:CI}
	\begin{alignat}{1}
	h(0,x) &= h_{\mathrm{stat}}(x), \\
	K(0,x) &= \frac{x}{2\eps}, \\
	\uu(0,x) &= 1/h_{\mathrm{stat}}(x), \label{eq:u_init}
	\end{alignat}
	\end{subequations}
where we approximate the initial condition \eqref{eq:u_init} on $\uu$ using the value resulting from equation \eqref{eq:mass_cons_ndim} in stationary regime. 

At first sight, the system of equations (\ref{eq:system_ndim}--\ref{eq:curvature_ndim}) should require five boundary conditions: two on the thickness $h$, two on the vertical velocity $\uu$ and one on the mean curvature $K$. However, the order of differentiation on $\uu$ can by reduced by one by defining $\lambda = \partial_x h$ as an intermediate variable. Using the mass conservation $h \partial_x \uu = -\partial_t h - \uu \lambda$ in equation \eqref{eq:tang_stress_ndim}, only first-order spatial derivative for the variables $h$, $\lambda$, $K$ and $\uu$ are left in the system. The four necessary boundary conditions are then imposed as follows. 

A position $x_0$ is fixed in the static meniscus, close to the surface of the liquid bath $x=0$ such that $0 < x_0 < L_0$, as sketched in figure~\ref{fig:notations}. The following dimensionless boundary conditions are set at that location:
	\begin{subequations}	\label{eq:CL_bath}
	\begin{alignat}{1}
	\partial_x h(t,x_0) &= h^{\prime}_{\mathrm{stat}}(x_0), \\
	K(t,x_0) &= \frac{x_0}{2\eps},
	\end{alignat}
	\end{subequations}
where the prime denotes the derivative with respect to the $x$-coordinate. The remaining boundary conditions are imposed on the fiber, moving upwards at a constant velocity $U$, and thus located at the dimensionless position $L(t) = L_0 + t$:
	\begin{subequations}	\label{eq:CL_fibre}
	\begin{alignat}{1}
	h(t,L(t)) &= 1, \\
	\uu(t,L(t)) &= 1. \label{eq:u_bc1}
	\end{alignat}
	\end{subequations}
We show in appendix~\ref{Appendix:x_0L_0} that the results are independent of the arbitrary position $x_0$ taken in the static meniscus and do not vary significantly with the initial position $L_0$ of the fiber, as long as $L_0 \leqslant \sqrt{2}$. These parameters will henceforth be set to $x_0 = 0.1$ and $L_0 = \sqrt{2}$, as justified in appendix~\ref{Appendix:x_0L_0}. 

The system of partial differential equations (\ref{eq:system_ndim}--\ref{eq:curvature_ndim}), supplemented by the initial solutions \eqref{eq:CI} and boundary conditions (\ref{eq:CL_bath}--\ref{eq:CL_fibre}), is solved using the direct solver MUMPS in COMSOL 5.0. Since the problem involves a moving boundary -- the fiber which is lifted at a constant velocity to create the film -- the domain geometry is changing with time. The mesh is thus deformed as prescribed by the Arbitrary Lagrangian Eulerian (ALE) algorithm.
%
\subsection{Static limit}
%
\begin{figure}
\centering
\includegraphics[width=10cm]{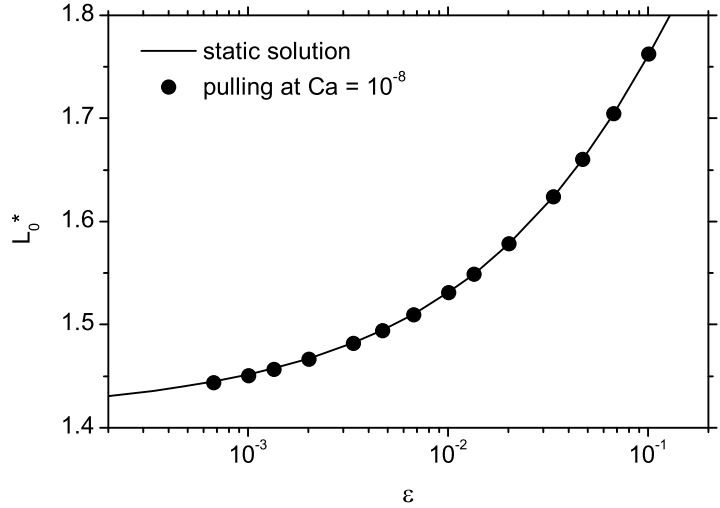}
\caption{The dimensionless break-up height $L_0^{\ast}$ of a static film, deduced from \eqref{eq:static_max_height}, is plotted as a function of the aspect ratio $\eps = H / \ell_c$ (solid line). The points correspond to the break-up height computed from our model when pulling at a capillary number $\Ca = 10^{-8}$, \textit{i.e.} in a quasi-static regime. The corresponding Weber number is $\We \approx 3 \times 10^{-18}$ and the Hamaker constant is set to $A_H=4.4\times 10^{-20}~\joule$.}
\label{fig:comp_static}
\end{figure}
We first check that the static limit is recovered for very small values of $\Ca$ in \eqref{eq:system_ndim}. Thanks to the analytical expression of the thickness profile \eqref{eq:static_meniscus_profile} in the static case, it can be shown \cite{PhD_Heller} that the point of minimal thickness in the static film is located at $x=\sqrt{2}$ for $\sqrt{2} < L_0 \le 2$. The (dimensionless) maximum height $L_0^{\ast}$ of a static film is then given by $h_{\mathrm{stat}}(\sqrt{2})=0$, hence the implicit analytical expression for $L_0^{\ast}$:
	\begin{equation}
        \eps = \sqrt{2} - \sqrt{4-(L_0^{\ast})^2} - \arctanh \!\left( \sqrt{2} \right)  + \arctanh \!\left( \frac{2}{\sqrt{4-(L_0^{\ast})^2}} \right).
	\label{eq:static_max_height}
	\end{equation}
In figure~\ref{fig:comp_static}, we compare the maximum height $L_0^{\ast}$ of a static film given by \eqref{eq:static_max_height} to the break-up height of a quasi-static film pulled at $\Ca = 10^{-8}$ using our model, for different fiber half-widths, \textit{i.e.} different values of $\eps$. The static limit is successfully recovered in our simulation.
%
\subsection{Comparison to the 2D case}
%
\begin{figure}
\begin{center}

\includegraphics[width=10cm]{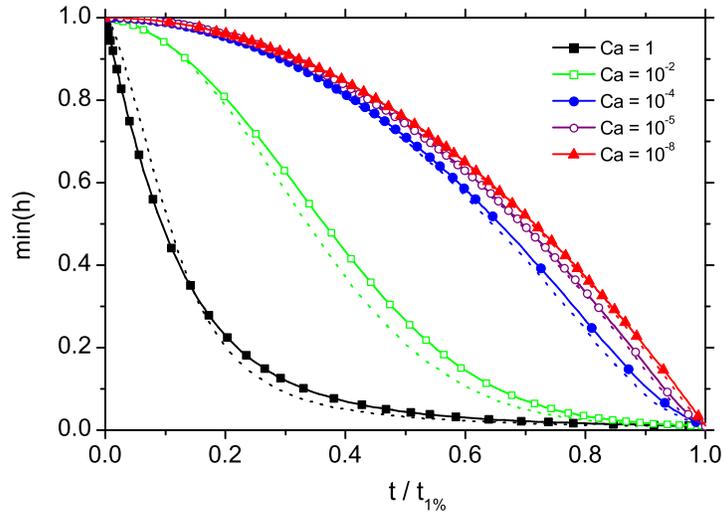}
\caption{The minimum film thickness $\mathrm{min}(h)$ is represented as a function of time $t$, normalised by the time $t_{1\%}$ when $\mathrm{min}(h)$ reaches $1\%$ of its initial value. For a given capillary number, the solid line with symbols is the drainage curve obtained from our lubrication approach, while the dotted line is the prediction of the 2D simulations reported by \cite{PhD_Heller}. In both cases, the aspect ratio is fixed to $\eps = 0.01$ and inertia and van der Waals forces are omitted ($\We=0$ and $A_H=0$)}.
\label{fig:comp_dyn_Heller}

\end{center}
\end{figure}
We now compare the results of our lubrication model in the dynamic regime to those reported in \cite{PhD_Heller}, who computed the Stokes equations in a two-dimensional and deformable domain. Note that both inertia and van der Waals forces were neglected in Heller's simulation, so we temporarily set $\We=0$ and $\mathcal{A}=0$ in \eqref{eq:system_ndim} for the sake of consistency. 

Figure~\ref{fig:comp_dyn_Heller} compares the drainage dynamics of films pulled at various velocities for a fixed aspect ratio $\eps = 0.01$, using our model (solid lines with symbols) and the 2D simulation results of \cite{PhD_Heller} (dotted lines). The minimum of the film thickness $\mathrm{min}(h)$ is plotted as a function of time, normalised by the time $t_{1\%}$ when the minimal thickness reaches 0.01, namely $1\%$ of its initial value. This normalisation allows to compare the drainage dynamics obtained for different values of the capillary number. 

The comparison presented in figure~\ref{fig:comp_dyn_Heller} shows that the lubrication approach is in good agreement with the corresponding 2D simulations, especially at small capillary numbers $\Ca \leqslant 10^{-4}$. A small discrepancy (less than $10\%$) is observed at capillary numbers $\Ca \geqslant 10^{-2}$, where the lubrication approximation predicts a slightly slower drainage than the 2D simulations, but the overall shape of the drainage curve is preserved. 
%
\section{Results of the model}  \label{sec:results_model}
%
In this section, we present the numerical results obtained when solving the model described in section \ref{sec:model}. In particular, we show that film rupture is a well-defined event, both in space and time. This allows us to introduce the critical thickness for rupture and the film lifetime (or equivalently the film break-up height), which we determine as functions of the aspect ratio $\eps$, the capillary number $\Ca$ and the dimensionless Hamaker constant $\mathcal{A}^{\prime}$.
%
\subsection{Thickness and velocity profiles} \label{sec:profils}
%
\begin{figure}
\centering
\includegraphics[width=\linewidth]{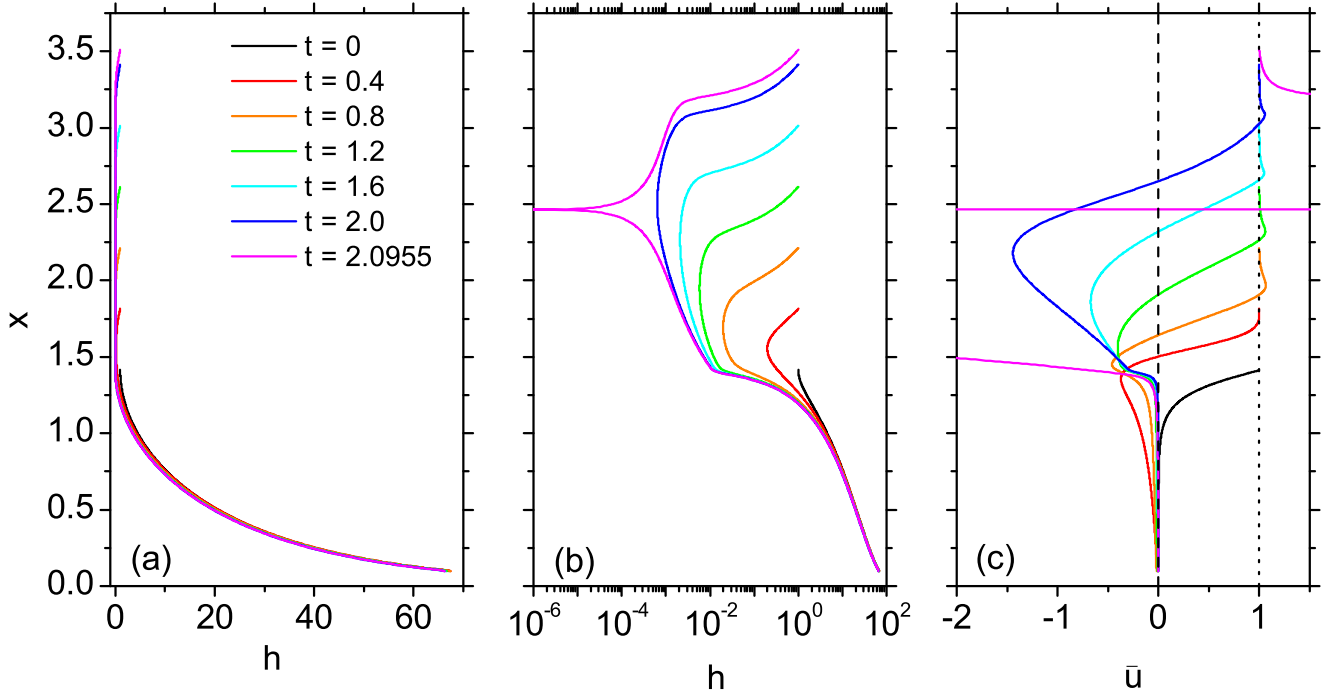}
\caption{Dimensionless thickness profiles $h(x)$ in (a) linear and (b) log scales and (c) average velocity profiles $\uu (x)$ at different times during film pulling. These profiles have been obtained by solving the model described in section~\ref{sec:model} with a pulling velocity $U=1~\mms$ ($\Ca=0.047$), a fiber half-width $H=50~\micro\meter$ ($\eps=0.034$) and a Hamaker constant $A_H=4.4\times 10^{-20}~\joule$. The film rupture, defined as the moment when the minimum dimensionless film thickness reaches $10^{-6}$, occurs at $t=2.0955$, corresponding in this case to approximately $3~\second$. The vertical dotted line at $\uu = 1$ corresponds to a (dimensional) average velocity equal to the pulling velocity $U$, while the dashed line represents $\uu =0$.}
\label{fig:profils}
\end{figure}
Typical film thickness profiles $h(x)$ are represented in figure~\ref{fig:profils} in (a) linear and (b) log scales, at different times during pulling. At $t=0$, a static meniscus spans between the liquid bath at $x=0$ and the fiber, which is initially located at $L_0 = \sqrt{2}$ and set into motion at a constant velocity $U$ ($U=1~\mms$ in figure~\ref{fig:profils}). As the fiber goes up, the liquid film grows in height and thins, as shown in figures \ref{fig:profils}a and \ref{fig:profils}b. Once the film has become sufficently thin, the van der Waals term $\mathcal{A} \, \partial_x h / h^3$ comes into play in equation \eqref{eq:tang_stress_ndim}, as we will see in a more quantitative way in section \ref{sec:hstar}. This term ultimately triggers the film break-up, which is characterised by a very rapid and localised drop of the film thickness, as demontrated by the magenta curve in figure \ref{fig:profils}b.

Figure \ref{fig:profils}c displays the spatial variation of the average velocity $\uu$ in the film at different times during film pulling, for the same parameters as in figures \ref{fig:profils}a and \ref{fig:profils}b. The initial velocity profile (black curve) differs slightly from the initial condition \eqref{eq:u_init} and is numerically converged towards to satisfy equations (\ref{eq:system_ndim} -- \ref{eq:curvature_ndim}). For $t>0.4$, the velocity in the static meniscus ($x\lesssim\sqrt{2}$) goes back to zero and the initial velocity profile is forgotten. At all times $t>0$, the average velocity $\uu$ is negative in the lower part of the film and positive in its upper part. Consequently, there is a location in the film where $\uu = 0$, which is close to the minimum in the thickness profile. The liquid is expelled from this minimum with a flow rate that seems to diverge at the point of rupture (magenta curve). The flow in the vicinity of the point of minimal thickness is further studied in appendix \ref{Appendix:streamlines}, where the streamlines are computed for various times during film pulling.
%
\subsection{Film drainage and rupture} \label{sec:drainage_rupture}
%
\begin{figure}
\centering
\includegraphics[width=10cm]{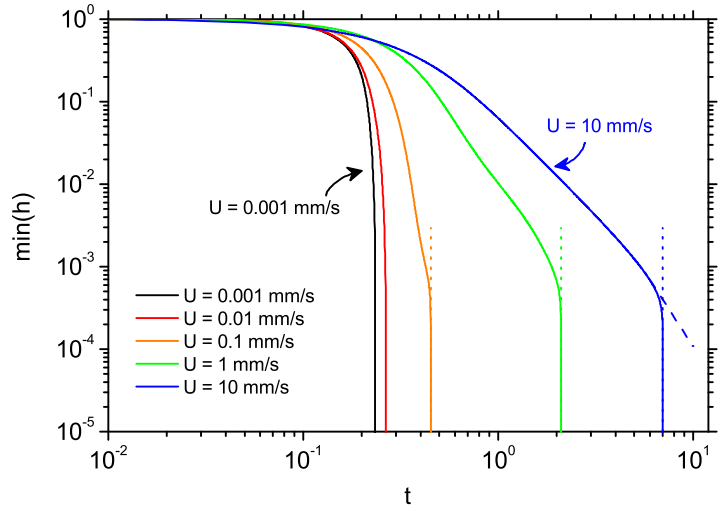}
\caption{The minimum thickness $\mathrm{min}(h)$ is plotted as a function of time in dimensionless units, for various values of the pulling velocity $U$ (solid lines) and a fixed fiber half-width $H=50~\micro\meter$ ($\eps=0.034$) and Hamaker constant $A_H=4.4\times 10^{-20}~\joule$. For $U \gtrsim 0.2~\mms$ ($\Ca \gtrsim 0.01$), a ``slow'' drainage regime is first observed, followed by a sharp decrease in thickness, corresponding to film rupture. The transition from the drainage to the rupture regime happens when the film reaches a critical thickness, which can be defined as the intersection between a power law fit of the end of the drainage regime (dashed line) and the rupture regime (dotted lines).}
\label{fig:mine_vs_time}
\end{figure}
We now follow the film thickness at the point where it is minimal as a function of time. The thinning dynamics at this point is displayed in figure~\ref{fig:mine_vs_time} (solid lines) for different values of the pulling velocity $U$. For velocities above $0.1~\mms$, the thinning dynamics exhibit two distinct parts: a drainage regime, where the thickness gently decreases with time, and a rupture regime, where the thickness drops abruptly. 

In practice, the simulation is stopped when the minimal thickness $\mathrm{min}(h)$ reaches $10^{-6}$, which corresponds to a subangstrom film thickness for the example displayed in figure \ref{fig:profils} (where $H = 50~\micro\meter$). The dimensionless film lifetime $\tau$ is then defined as the time when $\mathrm{min}(h) = 10^{-6}$ and the corresponding dimensionless break-up height is given by $\Lstar = L_0 + \tau = \sqrt{2} + \tau$. Note that this definition does not depend on the cut-off value $10^{-6}$ as long as the rupture regime -- where the slope is almost vertical -- has been reached. Thus, the break-up height of the film is an observable, which will be studied in details in paragraph \ref{sec:lmax}. Note that the model also gives a prediction for the location of the puncture in the film, which we compare to experimental observations in appendix \ref{Appendix:puncture_position}.
%
\subsection{Critical thickness for rupture} \label{sec:hstar}
%
\begin{figure}
\centering
\includegraphics[width=\linewidth]{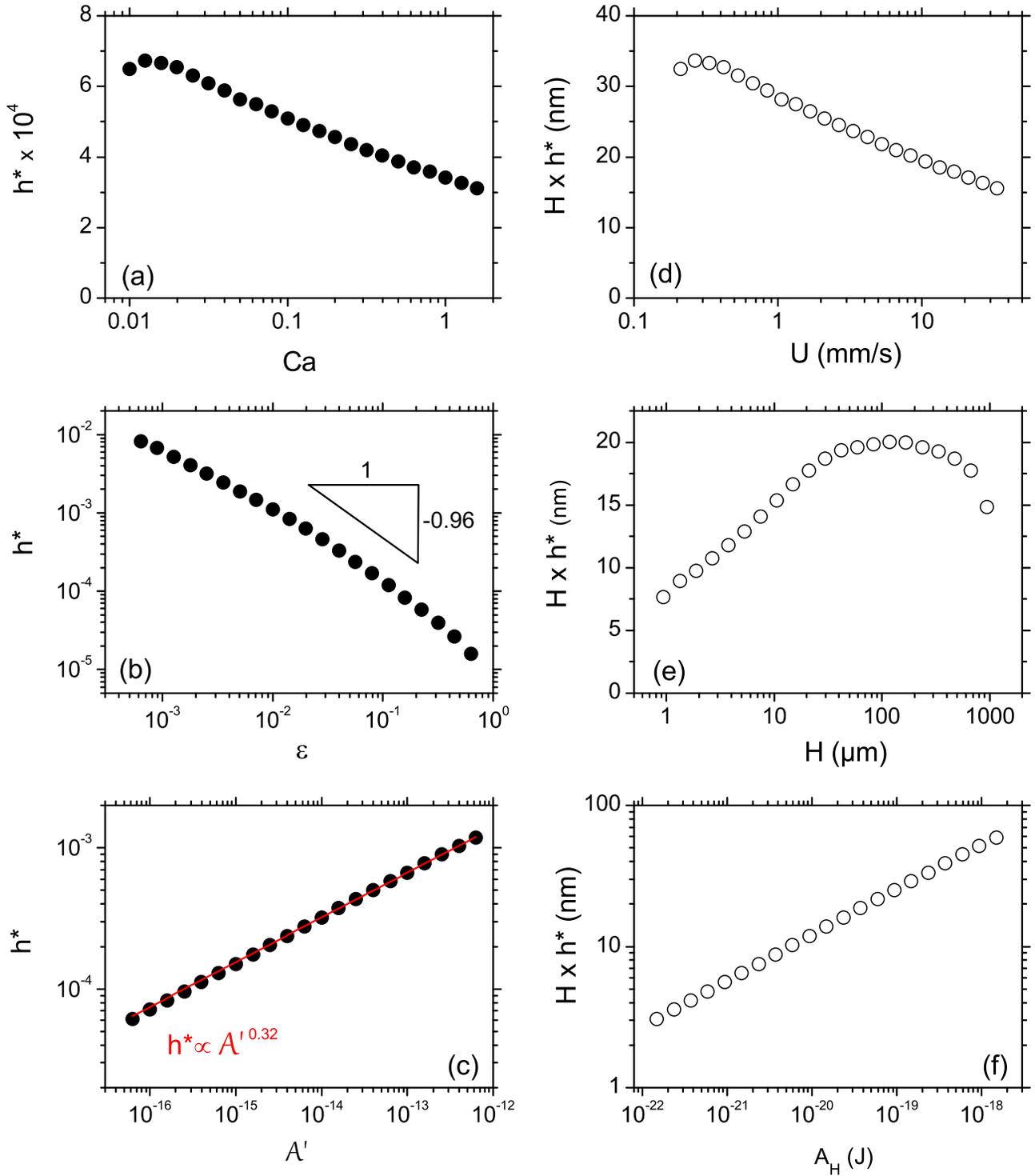}
\caption{The dimensionless (resp. dimensional) critical thickness for rupture $\hstar$ (resp. $H \times \hstar$) is plotted as a function of the control parameters, one being varied at a time with the others fixed to the following values: $\eps=0.034$ ($H=50~\micro\meter$), $\Ca = 0.47$ ($U=10~\mms$) and $\mathcal{A}^{\prime} = A_H / 16 \pi \gamma \ell_c^2 = 1.84 \times 10^{-14}$ ($A_H = 4.4 \times 10^{-20}~\joule$).}
\label{fig:hstar}
\end{figure}
The transition from the drainage regime to the rupture regime in figure~\ref{fig:mine_vs_time} occurs around a critical film thickness, when the van der Waals forces come into play. We define this critical thickness for rupture, denoted $\hstar$, as the intersection between a power law fit of the end of the drainage regime (dashed line in figure~\ref{fig:mine_vs_time}) and the rupture regime (dotted lines in figure~\ref{fig:mine_vs_time}). The variation of the critical thickness $\hstar$ with the capillary number $\Ca$, the aspect ratio $\eps$ and the dimensionless Hamaker constant $\mathcal{A}^{\prime} = A_H / 16 \pi \gamma \ell_c^2$ are respectively displayed in figures~\ref{fig:hstar}a, \ref{fig:hstar}b and \ref{fig:hstar}c. The dimensional counterparts of these data sets are shown in the right panel of figure \ref{fig:hstar}, where the dimensional critical thickness $H \times \hstar$ is plotted as a function of the pulling velocity $U$, the fiber half-width $H$ and the Hamaker constant $A_H$, respectively displayed in figures \ref{fig:hstar}d, \ref{fig:hstar}e and \ref{fig:hstar}f. \bigskip \\
\indent Figures \ref{fig:hstar}a and \ref{fig:hstar}d show that the critical thickness $\hstar$ decreases with the pulling velocity for capillary numbers in the range $0.01 \lesssim \Ca \lesssim 1$ where the drainage and rupture regimes are clearly distinguished (see figure \ref{fig:mine_vs_time}). For $\Ca \lesssim 0.01$, the critical thickness is no longer defined in our approach since the drainage and rupture regimes merge. The decrease of $\hstar$ with $\Ca$ shows that entrainment is a stabilising factor for the film, in the sense that the destabilising effect of van der Waals forces is delayed -- \textit{i.e.} comes into play for thinner films.

Figure \ref{fig:hstar}b indicates that the critical thickness also decreases with the aspect ratio $\eps$. This behaviour can be qualitatively understood from the fact that the scale of the van der Waals contribution in equation \eqref{eq:tang_stress_ndim} is set by the Hamaker number $\mathcal{A} = \mathcal{A}^{\prime} / \eps^3$. For a given dimensionless Hamaker constant $\mathcal{A}^{\prime}$, the larger the aspect ratio, the smaller the Hamaker number and so the smaller the thickness required for $\mathcal{A} \, \partial_x h / h^4$ to be of leading order, hence the decrease of $\hstar$ with $\eps$. More quantitatively, the variation of the critical thickness compares favourably with the power law $\hstar \propto \eps^{-0.96}$ for $\eps = 10^{-2}-10^{-1}$, as will be justified later. The critical thickness varies more slowly than $\eps^{-1}$ below $\eps \sim 10^{-1}$, but faster than $\eps^{-1}$ for $\eps \gtrsim 10^{-1}$, hence the non-monotonic behaviour of the dimensional critical thickness $H \times \hstar$ as a function of the fiber half-width $H$, as shown in figure \ref{fig:hstar}e. This change in behaviour for $\eps \gtrsim 10^{-1}$ is likely due to the $\eps^2$ term in the denominator of the mean curvature \eqref{eq:curvature_ndim}, which becomes non-negligible. \bigskip \\
\indent Finally, figures \ref{fig:hstar}c and \ref{fig:hstar}f display the variation of the critical thickness with the dimensionless Hamaker constant $\mathcal{A}^{\prime} = A_H / 16 \pi \gamma \ell_c^2$, which is a property of the liquid. The parameters $\Ca$ and $\eps$ remain fixed, while the Hamaker constant $A_H$ is varied in the range $10^{-22} - 10^{-18}~\joule$, all the other properties of the liquid being set to the values given in paragraph \ref{sec:ndim_problem}. The critical thickness $\hstar$ increases with the dimensionless Hamaker constant as a power law (red line in figure \ref{fig:hstar}c) with an exponent $\alpha \approx 0.32$. 

Such a power law behaviour for the critical thickness as a function of the Hamaker constant had already been predicted by \cite{Vrij1968} for small horizontal liquid films. Their approach consisted in calculating the film lifetime by summing the time needed for the film to reach a certain thickness and the time required for a van der Waals-driven instability to develop. Under the assumption of flat and rigid liquid/air interfaces, they derived the critical thickness, which was found to vary with the Hamaker constant as $A_H^{2/7}$ ($2/7 \approx 0.29$), an exponent which is close to our value of $0.32$. The difference probably stems from the flow in our continuously stretched film with stress-free interfaces, which is quite different from the Poiseuille flow expected for rigid interfaces.

We finally give arguments regarding how the power law behaviour of the critical thickness $\hstar$ \textit{vs} $\Aprime$ can be related to the variation of $\hstar$ with $\eps$. Figure \ref{fig:hstar}c shows that the critical thickness follows $\hstar = \beta {\Aprime}^{\alpha}$, where $\alpha$ and $\beta$ are \textit{a priori} functions of $\eps$ and $\Ca$ (for example, $\alpha = 0.32$ for $\eps=0.034$ and $\Ca=0.47$). Considering that film rupture is essentially driven by the Hamaker number $\mathcal{A} = \Aprime / \eps^3$, we can assume that, for a given capillary number $\Ca$, the critical thickness $\hstar$ is essentially a function of $\mathcal{A}$: $\hstar = f(\mathcal{A})$. Still for a fixed capillary number, this is equivalent to saying that $\hstar$ depends on $\eps$ only through $\mathcal{A} = \Aprime / \eps^3$, \textit{i.e.} $\hstar = f(\mathcal{A}) = f(\Aprime / \eps^3) = \beta(\eps) \times {\Aprime}^{\alpha(\eps)}$. This imposes that the exponent $\alpha$ is independent of $\eps$ and that $\beta(\eps) \propto \eps^{-3\alpha}$, \textit{i.e.} $\hstar \propto \eps^{-0.96}$ for $\Ca = 0.47$ and $\Aprime = 1.84 \times 10^{-14}$. Figure \ref{fig:hstar}b shows that the assumption that $\hstar = f(\mathcal{A})$ seems reasonable, but is not rigorously true for $\eps \lesssim 0.01$ and $\eps \gtrsim 0.1$, since the data deviate from the $\hstar \propto \eps^{-0.96}$ behaviour. \bigskip \\
\indent The variations of the critical thickness with $\Ca$ and $\eps$ point at the fact that $\hstar$ is not an intrinsic property of the liquid but also depends on the experimental conditions and flow dynamics. Numerous studies (see \textit{e.g.} \cite{Manev2005}) have been dedicated to the critical thickness of horizontal thin liquid films, from both the experimental and theoretical point of view. The results presented in figures \ref{fig:hstar}d and \ref{fig:hstar}e put into question the applicability of these static investigations to dynamic situations, such as foam generation for instance.
%
\subsection{Maximal height of the film} \label{sec:lmax}
%
\begin{figure}
\centering
\includegraphics[width=10cm]{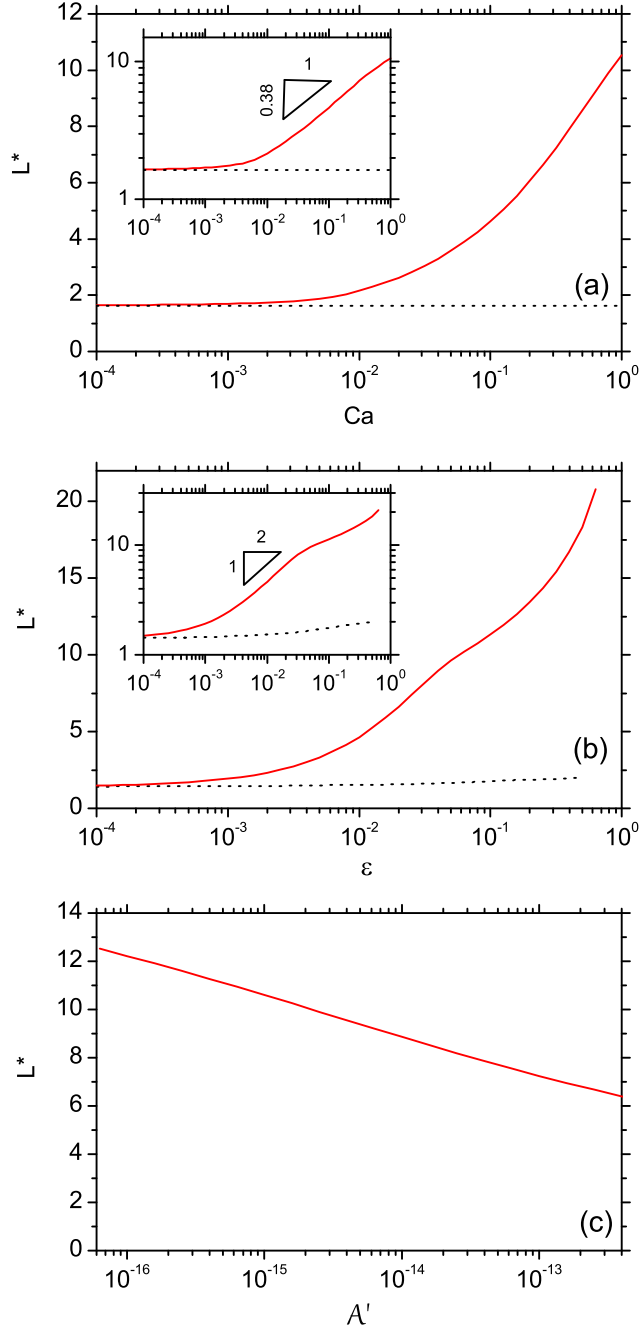}
\caption{The dimensionless film break-up height $\Lstar$ is plotted (a) as a function of the capillary number $\Ca$ for a fixed fiber half-width $H=50~\micro\meter$ ($\eps=0.034$) in log-linear and log-log scale (insert), (b) as a function of the aspect ratio $\eps$ for a fixed pulling velocity $U=10~\mms$ ($\Ca=0.47$) in log-linear and log-log scale (insert), and (c) as a function of the dimensionless Hamaker constant $\mathcal{A}^{\prime} = A_H / 16 \pi \gamma \ell_c^2$ for $H=50~\micro\meter$ and $U=10~\mms$. The dotted lines represent the static limit given by equation \eqref{eq:static_max_height}.}
\label{fig:lmax}
\end{figure}
As demonstrated in paragraphs \ref{sec:profils} and \ref{sec:drainage_rupture}, the height $\Lstar$ reached by the film when the rupture occurs is well-defined. In figure \ref{fig:lmax}, we report this break-up height as a function of the capillary number $\Ca$ (figure \ref{fig:lmax}a) and the aspect ratio $\eps$ (figure \ref{fig:lmax}b) for given liquid properties. For fixed pulling parameters, the effect of the dimensionless Hamaker constant $\mathcal{A}^{\prime}$ on the break-up height $\Lstar$ is displayed in figure \ref{fig:lmax}c. \bigskip \\
\indent For $\Ca \lesssim 10^{-3}$, the break-up height of the film is equal to the static limit given by equation \eqref{eq:static_max_height} (dotted line in figure \ref{fig:lmax}a). Beyond this value, $\Lstar$ increases with the capillary number, which is expected since the creation of the film is driven by viscous entrainement. This increase approximately follows a power law of exponent $0.38$, as shown in the insert in figure \ref{fig:lmax}a. Note that, for given liquid properties, the value of the exponent may \textit{a priori} depend on the aspect ratio $\eps$.

As displayed in figure \ref{fig:lmax}b, the break-up height $\Lstar$ is also an increasing function of the aspect ratio. This can be understood from the fact that the larger the aspect ratio $\eps$, the smaller the Hamaker number $\mathcal{A}= \mathcal{A}^{\prime} / \eps^3$ and the thinner the film when van der Waals forces come into play and trigger film rupture. The $\eps$ dependency of the break-up height seems to exibit three different regimes. For $\eps \lesssim 10^{-3}$, the static limit given by equation \eqref{eq:static_max_height} (dotted line in figure \ref{fig:lmax}b) is asymptotically reached. As shown in the insert in figure \ref{fig:lmax}b, the break-up height $\Lstar$ then appears to follow a power law increase of exponent $1/2$ up to $\eps \sim 3 \times 10^{-2}$. Beyond, it enters a third regime where the break-up height rises more slowly than $\eps^{1/2}$. This is likely again a signature of the $\eps^2$ term in the denominator of the mean curvature \eqref{eq:curvature_ndim}, which becomes non-negligible for $\eps \gtrsim 10^{-1}$.

Finally, the film break-up height decreases with the dimensionless Hamaker constant, due to the contribution of $\mathcal{A}^{\prime}$ to the Hamaker number $\mathcal{A}$. The break-up height turns out to be less sensitive to the Hamaker constant, varied in the range $A_H = 10^{-22} - 10^{-18}~\joule$, than to the fiber half-width $H$ or to the pulling velocity $U$. \bigskip \\
Note that solving the system without inertia -- \textit{i.e.} setting $\We = 0$ in equation \eqref{eq:tang_stress_ndim} -- essentially yields the same results. Although it plays virtually no part within the range of parameters explored here, inertia was kept in all simulations since it was found to help the convergence of the numerical solver used in COMSOL.
%
\section{Experiments: materials and methods  \label{sec:exp}}
%
\begin{figure}
\centering
\includegraphics[width=8cm]{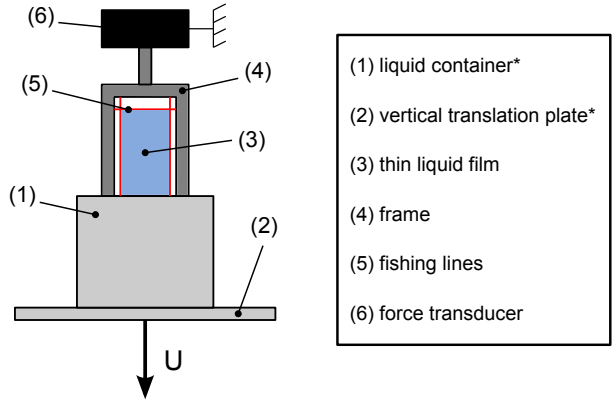}
\caption{Sketch of the experimental setup used to measure the maximal height of vertical liquid films. A container (1) filled with liquid is bound to a vertical motorised translation plate (2). By lowering the container at a constant velocity $U$, a thin liquid film (3) is deposited on a fixed frame (4) gridded with fishing lines (5). The break-up of this film is detected using the force transducer (6) to which the frame is attached. The stars indicate the elements of the setup which are mobile in the reference frame of the laboratory.}
\label{fig:setup}
\end{figure}
In the following, we measure experimentally the break-up height of pure liquid films in order to test the predictions of the model presented in section \ref{sec:results_model}. More specifically, we investigate the influence of the capillary number $\Ca$ by varying the pulling velocity $U$ and the liquid viscosity $\eta$, and the influence of the aspect ratio $\eps$ by using different fiber diameters.
%
\subsection{Experimental setup and protocole \label{protocole}}
%
The liquid films are formed on a $2 \times 9~\centi\meter$ rigid acrylonitrile butadiene styrene (ABS) frame covered in cross-linked polyepoxide to avoid liquid penetration. As sketched in figure~\ref{fig:setup}, the frame is gridded with one horizontal and two vertical nylon fishing lines of diameter $D=11, \, 57 \; \text{or} \; 175~\micro\meter$. The spacing between the vertical fishing lines is equal to $15~\milli\meter$. The frame is initially immersed in a liquid container, the horizontal fishing line being located at about $1~\milli\meter$ below the surface of the liquid bath. While the frame remains fixed during the whole experiment, the liquid container can be lowered at a constant velocity $0.01~\mms \leqslant U \leqslant 10~\mms$ using a motorised linear stage (Newport UTS 150 CC) coupled to a motion controller (Newport SMC 100 CC), hence creating a thin liquid film between the fishing lines. 

The film lifetime $\tau_{\mathrm{exp}}$ is defined in the experiments as the difference between the time when the horizontal fishing line crosses the surface of the liquid bath and the time when the film breaks. This experimental definition would correspond to an initial meniscus height $L_{0, \mathrm{dim}} = 0$ in the simulations. The film rupture is detected thanks to the force sensor (HBM, 5g) on which the frame is attached. When the film bursts, the signal measured by the force transducer undergoes a jump due to the disappearance of the surface tension contribution to the total vertical force exerted on the frame. When this jump is detected, the translation plate automatically stops and comes back to its initial position. This automated setup allows for statistical measurements of the film lifetime $\tau_{\mathrm{exp}}$ or, equivalently, of the film break-up height $\Lstar_{\mathrm{exp}} = U \tau_{\mathrm{exp}}$.
%
\subsection{Finding a pure liquid}
%

\begin{table*}
\centering
\begin{tabular}{ccccccc}
Property & Unit & $V350$ & $V1000$ & $V10~000$ & $V12~500$ & $V60~000$ \\
\hline
Viscosity $\eta$ & $\pascalsecond$ & 0.379 & 0.937 & 9.34 & 10.4 & 49.2 \\
Density $\rho$ & $\kilogrampercubicmetre$ & 970 & 970 & 970 & 973 & 973 \\
Surface tension $\gamma$ & $\mNparm$ & 21.2 & 21.3 & 21.1 & 21.2 & 21.1 \\
Molar mass $M_n$ & $\kilogram\per\mole$ & 16.5 & 25.7 & 58.3 & 60.2 & 90.1 \\
\end{tabular}
\caption{Main physical properties of the silicone oils used in the experiments. The viscosity and surface tension are given at $25~\celsius$ and have been measured after the fractionation processes for the $V350$ and $V1000$ oils. The experimental errors are $\lesssim 0.5 \%$ for viscosity measurements and $\pm 0.1~\mNparm$ for surface tension measurements. The number average molar mass $M_n$ is deduced from the viscosity using the expression $\log \eta = 1.00 + 0.0123 \, M_n^{1/2}$ \cite{Handbook}.}
\label{tab:PDMS}
\end{table*}
\begin{figure}
\centering
\includegraphics[width=10cm]{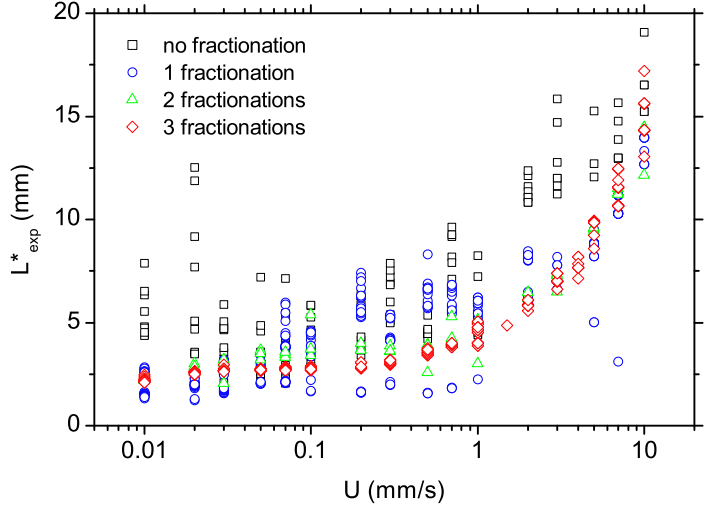}
\caption{The experimental break-up height $\Lstar_{\mathrm{exp}}$ of $V350$ silicone oil films is measured as a function of the pulling velocity $U$ with a nylon fiber of diameter $D = 175~\micro\meter$. The various symbols/colours show the data obtained after the application of $n$ fractionations ($0 \leqslant n \leqslant 3$) to the silicone oil taken from the bottle. The fractionation protocole, which aims at removing the shortest PDMS chains, significantly improves the reproducibility of the measurements.}
\label{fig:fractionation}
\end{figure}
In our experiments, the liquid used to form the films is silicone oil (methyl-terminated polydimethylsiloxane, PDMS, purchased from either Sigma Aldrich or Bluestar silicones), which is often considered as a model Newtonian pure liquid. The physicochemical parameters of the different silicone oils tested in the experiments are summed up in table~\ref{tab:PDMS}. The viscosities and surface tensions were respectively measured with a Physica MCR 300 rheometer and the rising bubble technique (Tracker, TECLIS), in the temperature range $T = 18-25~\celsius$. The Hamaker constant is expected to be the same for all the silicone oils used in this study and equal to $A_H = 4.4 \times 10^{-20}~\joule$ at $20~\celsius$ \cite{Drummond1997}.

As the liquid viscosity is varied over more than two orders of magnitude, the other physicochemical parameters -- \textit{i.e.} density and surface tension (see table \ref{tab:PDMS}) as well as the Hamaker constant -- remain virtually constant. This allows to consider that, for a given pulling velocity $U$ and a given fiber half-width $H$, changing the liquid will only affect the viscosity, hence the value of the capillary number $\Ca$, while leaving the other dimensionless parameters $\eps$, $\We$ and $\mathcal{A}$ unchanged.\bigskip \\
\indent The model we have built in section \ref{sec:model} is valid only for pure liquids, \textit{i.e.} in the absence of tangential stress at the interfaces. The experimental difficulty is that any small surface tension gradient -- due to temperature inhomogeneity or surface active agents for example -- will invalidate the stress-free boundary condition at the interfaces (equation \eqref{eq:surf_balance_T}). Because of their high surface tension, water surfaces are particularly prone to be contaminated by surface active impurities, which can lead to surface tension gradients. It has been shown that the drainage of a water film between two bubbles can be modelled using a no-slip boundary condition for surface tension gradients as small as $0.1~\mNparm^2$ \cite{Yaminsky2010}. This is the reason why we turned to silicone oil as a model pure liquid in the first place.

Commercially available silicone oils are actually mixtures of PDMS chains of different lengths (or equivalently of different molar masses), the distribution of which is peaked around the number average molar mass $M_n$. The surface tension of silicone oil increases with the number average molar mass up to $M_n \sim 10^4~\gram\per\mole$, where the surface tension saturates at a constant value $\gamma \approx 21~\mNparm$ \cite{Handbook}. Since short PDMS chains (\textit{i.e.} with a molar mass less than about $10^4~\gram\per\mole$) have a lower surface tension than longer chains, any spatial heterogeneity in their concentration -- either due to the creation of fresh interface or to evaporation -- can generate surface tension gradients. This phenomenon was already observed by \cite{Bascom1964} to have a strong influence on the spreading of oil on solid surfaces. Similarly, we suspect that traces of short PDMS chains may be sufficient to create interfacial shear stress in our silicone oil films, thus resulting in their overstabilisation.

\indent To avoid this problem, we use a basic fractionation protocole to get rid of the shortest chains in silicone oil. The commercial silicone oil is poured into a separatory funnel with about three times as much acetone (in volume). The mixture is vigorously shaken for five minutes and left to decant for several hours. Short PDMS chains are soluble in acetone while longer chains are not, so when the phase separation is complete, the PDMS phase contains less short chains than initially. The PDMS phase is then extracted, left in an oven at $60~\celsius$ for at least $20~\hour$ and finally put in a vacuum chamber for at least $20~\hour$ to remove the remaining acetone. 

Figure \ref{fig:fractionation} shows the break-up height of $V350$ silicone oil films as a function of the pulling velocity for a non-fractionated and (up to three times) fractionated oils. While the data corresponding to the non-fractionated oil is scattered, the application of the fractionation protocole described above yields much more reproducible results. Moreover, fractionated oils lead, on average, to shorter films than the non-fractionated oil, supporting the hypothesis of surface tension gradients due to short PDMS chains. The lower the viscosity, the more short chains in the oil. Thus, the fractionation protocole had to be applied three times to the $V350$ oil, but only once to the $V1000$ oil to get reproducible results. No fractionation was needed for the most viscous oils ($V10000$, $V12500$ and $V60000$). 
\vspace{-1mm}
%
\section{Experiments: results and comparison to theory \label{sec:comparison}}
%
\subsection{Influence of the pulling velocity}
%
\begin{figure}
\centering
\includegraphics[width=10cm]{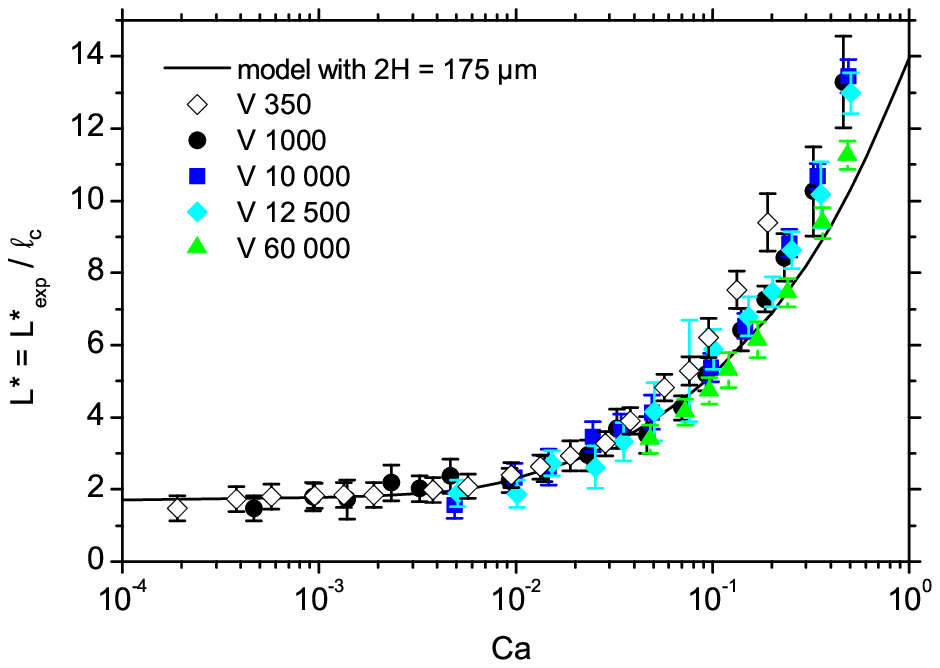}
\caption{The break-up height $\Lstar_{\mathrm{exp}}$ of films made from silicone oils of various viscosities is measured as a function of the pulling velocity $U$, using the same nylon fiber of diameter $D=175~\micro\meter$. The film break-up height is non-dimensionalised by the capillary length $\ell_c = 1.49~\milli\meter$ and displayed as a function of the capillary number $\Ca = \eta \, U / \gamma$. The non-dimensionalised experimental data (symbols) collapse onto a single mastercurve, which is compared to the prediction of our model (solid line, see also figure \ref{fig:lmax}a) with $2H = 175~\micro\meter$.}
\label{fig:comp_viscosity}
\vspace{1cm}
\includegraphics[width=10cm]{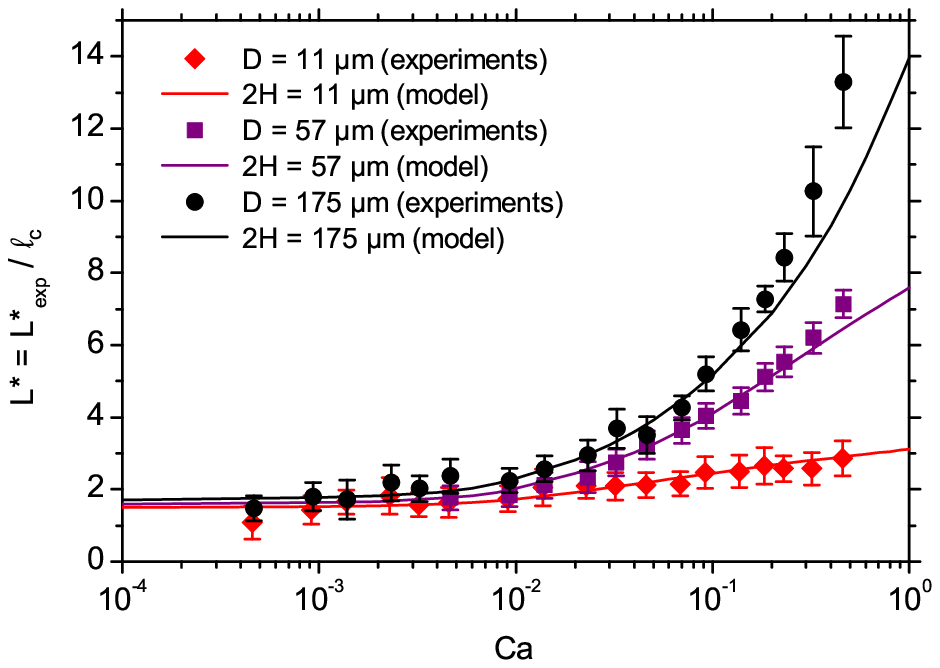}
\caption{The break-up height $\Lstar_{\mathrm{exp}}$ of films made from silicone oil is measured as a function of the pulling velocity $U$, using nylon fibers of various diameters. The experimental data (symbols) is non-dimensionalised as in figure~\ref{fig:comp_viscosity} and compared to the predictions of our model (solid lines), where $2H$ has been set equal to the fiber diameter $D$. The data for $D=11$ and $175~\micro\meter$ were obtained using the $V1000$ silicone oil, while the experiments for $D=57~\micro\meter$ were carried out with the $V10~000$ silicone oil.}
\label{fig:comp_fibre_diameter}
\end{figure}
Using a nylon fiber of fixed diameter $D = 175~\micro\meter$, the influence of the pulling velocity $U$ on the film break-up height is first explored for silicone oils of various viscosities. The film break-up height is measured using the protocole described in section~\ref{protocole} and the experiment is repeated at least twenty times for a given velocity and a given liquid viscosity. In the results presented in figure~\ref{fig:comp_viscosity}, each point is thus an average over at least twenty points and the error bar represents the square root of the quadratic sum of the standard deviation and a systematic error of $\pm 0.5~\milli\meter$ on the position of the liquid bath surface. It can be noted that the standard deviation of the measurements increases with $\Ca$, which may be due to the finite acquisition frequency of the force sensor used to detect film rupture.

Data points were obtained with pulling velocities varying over three orders of magnitude and liquid viscosities varying over two orders of magnitude. They collapse onto a single master-curve when plotted versus the capillary number $\Ca = \eta \, U / \gamma$. Note that the data corresponding to the less viscous oil ($V350$) lie slightly above the others for $\Ca \gtrsim 0.1$, perhaps because of residual surface tension gradients due to short PDMS chains that were not removed by the fractionation protocole. 

The experimental data is in good agreement with our model with $2H = 175~\micro\meter$ (solid line in figure~\ref{fig:comp_viscosity}) up to $\Ca \sim 0.1$. A small deviation from the theoretical prediction is observed at capillary numbers larger than $0.1$, where the films last longer than expected from the model, as will be discussed in paragraph \ref{sec:discussion}.
%
\subsection{Influence of the fiber diameter}
%
The effect of the diameter $D$ of the horizontal fiber which supports the film is then investigated experimentally. The film break-up height is measured as a function of the capillary number for three different fiber diameters, $D=11, \, 57 \; \text{and} \; 175~\micro\meter$  and the results are shown in figure~\ref{fig:comp_fibre_diameter}. As in figure~\ref{fig:comp_viscosity}, each data point is averaged over at least twenty measurements. The theoretical predictions computed from our model, setting $2H=11, \, 57 \; \text{and} \; 175~\micro\meter$ (\textit{i.e.} $\eps=0.0037$, $0.019$ and $0.059$, respectively), are also displayed in figure~\ref{fig:comp_fibre_diameter}.

For all fiber diameters, the experimental data are in good agreement with the theoretical predictions for $\Ca \leqslant 0.1$. The deviation previously observed at higher capillary numbers for the thickest fiber ($D=175~\micro\meter$) is reduced as the fiber diameter diminishes and the data for the thinnest fibers ($D=57~\micro\meter$ and $D=11~\micro\meter$) stay in line with the model for the whole range of capillary numbers probed in the experiments.
%
\subsection{Discussion} \label{sec:discussion}
%
%
\begin{figure}
\centering
\includegraphics[width=13cm]{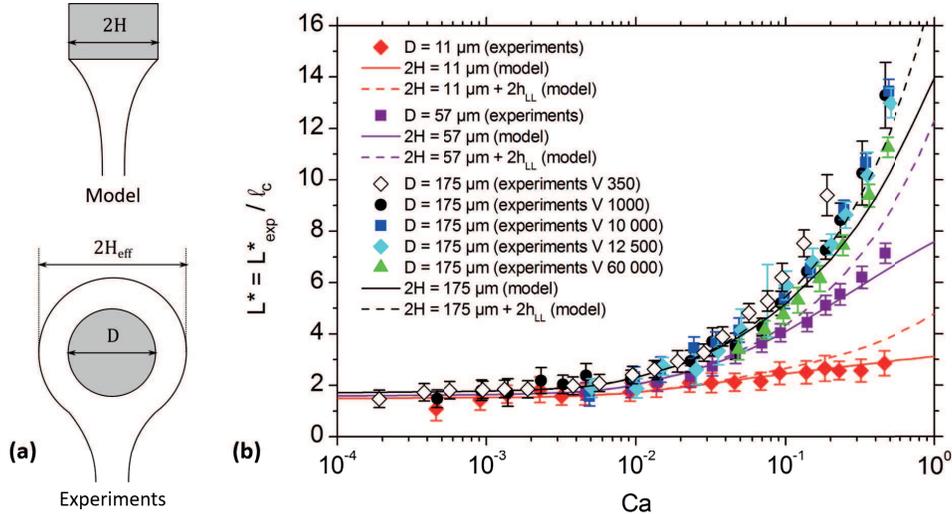}
\caption{(a) Sketches of the boundary condition at the fiber in the model (top), assuming a rectangular cross-section of width $2H$, and in the experiments (bottom), where the fiber has a circular cross-section of diameter $D$, certainly coated with a liquid layer, leading to an effective fiber width $2\Heff > D$. (b) The sets of experimental data displayed in figures~\ref{fig:comp_viscosity} and~\ref{fig:comp_fibre_diameter} are replotted together (symbols) and compared to the predictions of the model (dashed lines) using an effective fibre $2\Heff = D + 2h_{\mathrm{LL}}$ to account for the coating of a liquid layer of thickness $h_{\mathrm{LL}}$ around the fibre. The solid lines are again the predictions of the model assuming $2H=D$.}
\label{fig:CL_fibre}
\end{figure}
The simulations are in quantitative agreement with the experimental data, except at capillary numbers $\Ca \gtrsim 0.1$, where the maximal film length is actually underestimated by the simulations. Several hypotheses can be proposed to explain this discrepency. First, we showed that the results of the full-2D approach \cite{PhD_Heller} are successfully recovered by our lubrication model for $\Ca <  10^{-2}$, but that a small deviation appears for larger capillary numbers (see figure \ref{fig:comp_dyn_Heller}). The difference between our model and the experimental break-up height observed for $\Ca \gtrsim 0.1$ may then be attributed to 2D effects that are not captured in our lubrication analysis. We can also suspect that minute quantities of short PDMS chains may have escaped the fractionation process and could be sufficient, at high capillary numbers, to overstabilise the films.

Since the deviation from the theoretical prediction depends on the fiber diameter, we may also seek its explanation in the boundary condition at the top of the film, where it meets the supporting fiber. In the model, we impose that the thickness $2h$ at the top of the film is equal to the width $2H$ of the fiber. To compute the solid lines in figures \ref{fig:comp_viscosity} and \ref{fig:comp_fibre_diameter}, we assumed that $2H$ is equal to the diameter $D$ of the nylon fishing line used in the experiments. However, in practice, we can expect the nylon fiber to be coated with a liquid layer as it crosses the surface of the liquid pool, thus corresponding to an effective boundary condition with $2\Heff > D$, as pictured in figure \ref{fig:CL_fibre}a. Since the film break-up height $\Lstar$ is an increasing function of $\eps$, \textit{i.e.} of $H$, see figure \ref{fig:lmax}b, the underestimation of $H$ can lead to an underestimation of $\Lstar$.

In order to get a more quantitative insight into the effect of fiber coating on the film break-up height, we need to estimate the thickness of the liquid layer coated onto the horizontal fiber as it crosses the interface. This is actually a complex time dependent problem, which is however related to classical problem of \emph{vertical} fiber coating. The latter was solved in stationary regime by \cite{Landau1942}, who found an entrained thickness $h_{\mathrm{LL}}$ proportional to $\Ca^{2/3}$. Assuming that $h_{\mathrm{LL}}$ gives the correct order of magnitude for the thickness of the liquid layer coated onto a \emph{horizontal} fiber, we have solved our model (section \ref{sec:model}) where $2H$ was replaced by the effective value
\begin{equation}
2\Heff (\Ca) = D + 2h_{\mathrm{LL}} (\Ca) = D\left(1 + 1.34 \, \Ca^{2/3} \right).
\label{eq:enrobage}
\end{equation}
The corresponding predictions are dispayed as dashed lines in figure \ref{fig:CL_fibre}b, along with the experimental data of figures \ref{fig:comp_viscosity} and \ref{fig:comp_fibre_diameter} (symbols), and with the predictions of the model assuming $2H=D$ (solid lines, same as in figure \ref{fig:comp_fibre_diameter}). As expected from equation \eqref{eq:enrobage}, the influence of the coated layer becomes significant only for $\Ca \gtrsim 0.1$. It leads to a better agreement between the model's prediction and the experimental data for the thickest fiber ($D = 175~\micro\meter$), but to an overestimation of the film break-up height for the thinnest fibers ($D = 11~\micro\meter$ and $D = 57~\micro\meter$).

This overestimation of the break-up height is not particularly surprising since our way of taking into account the coated liquid layer is quite crude. The coated layer is not simply static during film pulling but can also drain by capillarity with a typical time of order $\eta D/2\gamma$ \cite{Kocarkova2013}, which is at most $0.3~\second$ for the thickest fiber ($D=175~\micro\meter$) and the most viscous oil ($V60000$). This drainage time is always much shorter than the experimental film lifetime, which lies in the range $\tau \sim 2 - 240~\second$, depending on the pulling velocity. The liquid layer coated onto the fiber may then feed the film and thus delay its rupture in a way which is not straightforward to estimate. Finally, above arguments show that the discrepancy between the presented model and the experiments at high capillary number may be explained by fiber coating. However, taking this effect into account in the simulation a more quantitative way would require the development of a model for horizontal fiber coating and thus lies beyond the scope of this paper.
%
%
\section{Conclusions \label{sec:conclusions}}
%
In this article, we described the pulling of a liquid film with stress-free interfaces by the means of a non-stationary model in the lubrication approximation. The entire life of the film could be addressed, from an initial static meniscus bridging the gap between a horizontal fiber and a liquid pool, until film rupture under the action of van der Waals forces. The film rupture turned out to be well-defined by a swift and localised drop of the film thickness, allowing the definition of the critical thickness for rupture $\hstar$ for capillary numbers $\Ca \gtrsim 0.01$ and of the film break-up height $\Lstar$. Both quantities were computed as functions of the capillary number $\Ca$, the aspect ratio $\eps$ and the dimensionless Hamaker constant $\mathcal{A}^{\prime}$.

Experiments were performed on thin films made of silicone oils of various viscosities and for three different fiber diameters. The predictions of the model turned out to be in very good agreement with the film break-up heights measured experimentally for fiber diameters varying over more than one order of magnitude and capillary numbers varying over three orders of magnitude. A small deviation was observed for $\Ca \gtrsim 0.1$ for the thickest fiber, which may be qualitatively explained by the coating of a liquid layer around the fiber as it crosses the surface of the liquid pool. 

We also found that the experimental break-up heights were reproducible only if the less viscous silicone oils were fractionated in order to remove the shortest PDMS chains, which were suspected to give rise to surface tension gradients during film pulling. This emphasises how sensitive film pulling expriments can be to minute quantities of surface active impurities and thus how strong the hypothesis of stress-free interfaces is.
%
%
\appendix
\section{Influence of $x_0$ and $L_0$ on the film break-up height} \label{Appendix:x_0L_0}
%
\begin{figure}
\centering
\includegraphics[width=12cm]{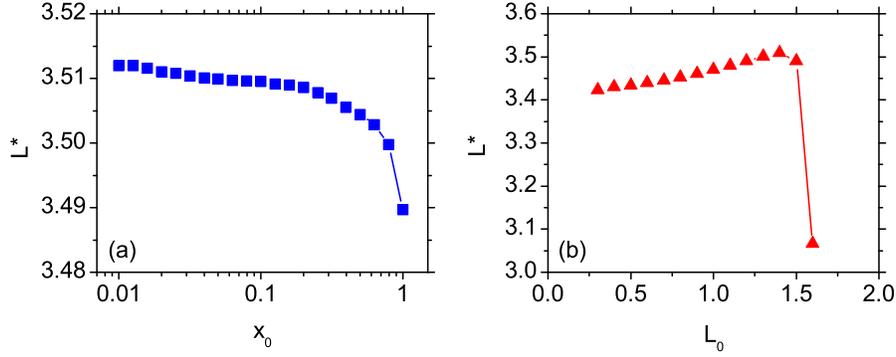}
\caption{Predictions of the model for the influence of (a) the position $x_0$ in the static meniscus at which the boundary conditions are imposed ($L_0=\sqrt{2}$ being fixed) and (b) initial height of the static meniscus $L_0$ ($x_0 = 0.1$ being fixed). The pulling velocity and fiber half-width are $U=1~\mms$ ($\Ca = 0.047$) and $H = 50~\micro\meter$ ($\eps = 0.034$) respectively.}
\label{fig:lmax_vs_x0_and_L0}
\end{figure}
In this appendix, we check that our main observable, namely the film break-up height $\Lstar$, is reasonably independent of the arbitrary position $x_0$ where the boundary conditions close to the liquid bath are set and of the initial height $L_0$ of the static meniscus. Figure \ref{fig:lmax_vs_x0_and_L0}a shows  that the break-up height does not vary by more than $0.2\%$ within the range $0.01 \leqslant x_0 \leqslant 0.5$. We have chosen to use the intermediate value $x_0 = 0.1$ in order to avoid the large thickness gradients in the vicinity of the liquid bath (arising when $x_0$ is too close to zero), while keeping a reasonably large initial domain (of length $L_0-x_0$).

Similarly figure \ref{fig:lmax_vs_x0_and_L0}b indicates that the initial position $L_0$ of the fiber does not change the value of $\Lstar$ by more than $2\%$ as long as $L_0$ is large enough compared to $x_0$ and less than $\sqrt{2}$. In the simulations, the initial position of the fiber has thus been set to the maximal value within this range, namely $L_0 = \sqrt{2}$, in order to have the largest possible initial domain and therefore the most mesh elements, allowing to keep a good spatial resolution when the mesh is stretched during film pulling. Note that for $L_0 > \sqrt{2}$, the thickness profile of the initial static meniscus features a minimum \cite{PhD_Heller}, which may explain the drop in the film break-up height observed for initial film heights greater than $\sqrt{2}$.
%
%
\section{Streamlines in the vicinity of the rupture point} \label{Appendix:streamlines}
\begin{figure}
\centering
\includegraphics[width=\linewidth]{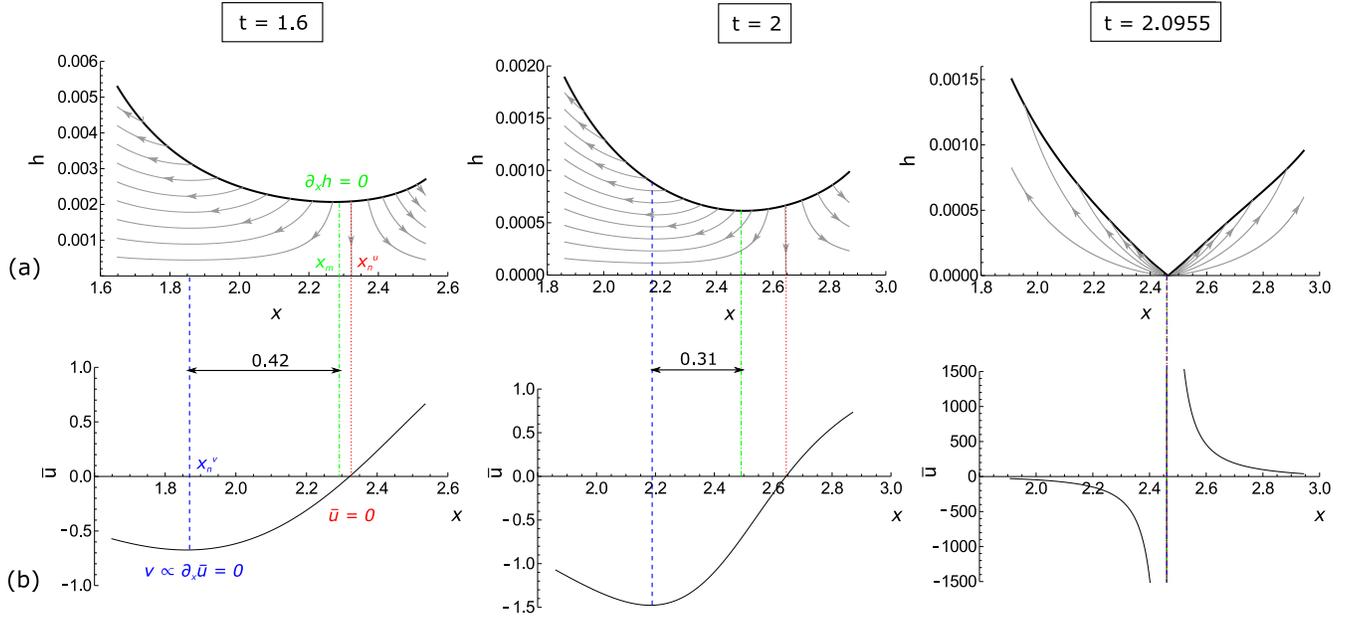}
\caption{Morphology of the flow in a film generated with a pulling velocity $U=1~\mms$ ($\Ca=0.047$), a fiber half-width $H=50~\micro\meter$ ($\eps=0.034$) and a Hamaker constant $A_H=4.4\times 10^{-20}~\joule$ (same parameters as in figure \ref{fig:profils}). For three different times, we display (a) the streamlines (thin black lines) superimposed on the film thickness profiles (thick black lines) and (b) the average velocity profiles $\bar{u}(x)$.}
\label{fig:streamlines}
\end{figure}

%
In this appendix, we compute the streamlines in order to visualise the flow in the vicinity of the rupture point. Figure \ref{fig:streamlines}a shows the streamlines for three different times -- $t=1.6$, $t=2$ and $t=2.0955$, namely at rupture -- during the pulling of a film with the same parameters as in figure \ref{fig:profils}. They were obtained by plotting the iso-contours of the (dimensionless) stream function
\begin{equation}
\psi = \int u \dd y = \bar{u}y + \eps^2 \int u_1 \dd y.
\label{eq:stream_function}
\end{equation}
The contribution of $u_1$ to the stream function $\psi$ was found to affect only marginally the streamlines ; we thus used the approximation $\psi \approx \bar{u}y$ in figure \ref{fig:streamlines}a. The corresponding average velocity profiles $\bar{u}(x)$ are displayed in figure \ref{fig:streamlines}b.

Several remarkable positions can be introduced:
\begin{itemize}
\item the point of minimum film thickness (green dash-dotted line in figure \ref{fig:streamlines}), located at a vertical position $x_{\mathrm{m}}$ and defined by $\partial_x h = 0$;
\item the point where the vertical velocity $\bar{u}$ changes sign (red dotted line in figure \ref{fig:streamlines}), located at $x_{\mathrm{n}}^u$ and defined by $\bar{u} = 0$;
\item the point where the horizontal velocity $v$ changes sign (blue dashed in figure \ref{fig:streamlines}), located at $x_{\mathrm{n}}^v$ and defined by $\partial_x \bar{u} = 0$ according to equation \eqref{eq:v_approx}.
\end{itemize}
One can see from figure \ref{fig:streamlines} that the minimum of film thickness $x_{\mathrm{m}}$ is initially located close to the point of zero vertical velocity $x_{\mathrm{n}}^u$. As time goes by and the film is stretched, $x_{\mathrm{m}}$ seems to be moving away from $x_{\mathrm{n}}^u$ and getting closer to the point of zero horizontal velocity $x_{\mathrm{n}}^v$. All three points merge at the approach of film rupture ($t=2.0955$).
%
\section{Location of the puncture in the film} \label{Appendix:puncture_position}
%
In addition to predicting the break-up height of the film, the model developed in section \ref{sec:model} also gives a prediction for the location of the puncture in the film (see \textit{e.g.} figure \ref{fig:profils}b), which calls for experimental verification. In this appendix, we sum up our observations regarding the bursting of silicone oil films, recorded with a high-speed camera (Photron Fastcam SA3) at frame rates ranging from $10~000$ to $25~000$ fps. We tested films made of $V350$ (pulling velocities $2-10~\mms$), $V1000$ (pulling velocities $1-5~\mms$) and $V10~000$ (pulling velocities $0.3-1~\mms$) silicone oils.

\begin{figure}
\centering
\includegraphics[width=10cm]{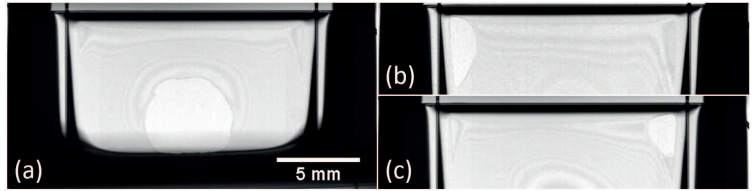}
\caption{High-speed observations of the bursting of films generated from a $V350$ silicone oil at a pulling velocity $U=10~\mms$. The location of the puncture was not reproducible from one film to another and the films were observed to rupture either (a) at the bottom center or (b) in the upper left corner or (c) in the upper right corner.}
\label{fig:position_rupture}
\end{figure}

Our first observation was that the film thickness is not homogeneous in the horizontal direction ($z$-axis), as can be seen qualitatively from the interference fringes in figure \ref{fig:position_rupture}. This may be due to boundary effects caused by the finite width of the film.

Some rupture events occur in the center of the film (with respect to the $z$-direction), as exemplified in figure \ref{fig:position_rupture}a. This is expected from our model, which assumes the film to be invariant along the $z$-axis. Under the same conditions, some rupture events are also observed in the upper corners of the film, as shown in figures \ref{fig:position_rupture}b and c. However, the location of the puncture does not seem to affect the overall film lifetime, which was shown to be reproducible for a given capillary number (see figure \ref{fig:comp_viscosity}). Thus, no matter where the film punctures, it always bursts at the same time, meaning that the thickness of the thin zones in the center and upper corners of the film follows roughly the same time evolution.
%
%
%
\section{acknowledgments}
The authors are grateful to Wiebke Drenckhan and Javier Rivero for fruitful discussions, and to Tycho van Noorden and Frank de Pont from COMSOL company for their great support. Many thanks also to Thibaut Gaillard and Delphine Hannoy for helping with the experiments. L.C. was supported by ANR F2F. B.S. thanks the F.R.S.-FNRS and the IAP-7/38 MicroMAST project for supporting this research. This work was performed under the umbrella of COST Action MP1106.

%
%
\bibliography{biblio_Benoit} 
\bibliographystyle{unsrt}
%
%
\end{document}